\documentclass[twocolumn]{aa} 

\usepackage[dvipsnames]{xcolor}

\usepackage{tabularx}
\usepackage{array}

\newcolumntype{Y}{>{\centering\arraybackslash}X}
\usepackage{graphicx}
\usepackage{txfonts}
\usepackage{hyperref}

\begin{document} 

\newcommand{\afffias}{Frankfurt Institute for Advanced Studies (FIAS), Ruth-Moufang-Strasse~1, 60438 Frankfurt am Main, Germany}
\newcommand{\affbgu}{Physics Department, Ben-Gurion University of the Negev, Beer-Sheva 84105, Israel}
\newcommand{\affbul}{Institute for Nuclear Research and Nuclear Energy, Bulgarian Academy of Sciences, Sofia, Bulgaria}

   \title{ Constraining the dark energy models using  {Baryon Acoustic Oscillations}: An approach independent of $H_0 \cdot r_d$}

   \author{Denitsa Staicova\inst{1} \and  David Benisty\inst{2,3}}

   \institute{
$^{1}\,$ Institute for Nuclear Research and Nuclear Energy, Bulgarian Academy of Sciences, Sofia, Bulgaria 
\\
$^{2}\,$ DAMTP, Centre for Mathematical Sciences, University of Cambridge, Wilberforce Road, Cambridge CB3 0WA, UK 
\\$^{3}\,$ Kavli Institute of Cosmology (KICC), University of Cambridge, Madingley Road, Cambridge, CB3 0HA, UK}
   \date{}

 
  \abstract{  The  {$H_0$ tension and the accompanying $r_d$} tension are a hot topic in current cosmology. In order to remove the degeneracy between the Hubble parameter $H_0$ and the sound horizon scale $r_d$ from the Baryon Acoustic Oscillations (BAO) datasets, we redefine the likelihood by marginalizing over the $H_0\cdot r_d$ parameter and then we perform full Bayesian analysis for different models of dark energy (DE). We find that our uncalibrated by early or late physics  {datasets cannot} constrain the DE models properly without further assumptions. By adding the type IA supernova  {dataset}, the models are constrained better with smaller errors on the DE parameters. The  {two} BAO datasets  {we use -- one with angular measurements and one with angular and radial ones} with their covariances, show statistical preferences for different models, with $\Lambda$CDM being the best model for  {one of them}.  Adding the Pantheon SnIA dataset with its covariance matrix boosts the statistical preference for $\Lambda$CDM.   
}
  


%

\maketitle
\section{Introduction}

 {A turning point in modern cosmology is the measurement of } the Hubble constant $H_0$  {revealing} the current accelerated expansion of the Universe~\cite{Riess:1998cb,Freedman:2010xv}.  Estimation of $H_0$ from the late Universe  {can be obtained from direct measurements such as distance ladders, strong lensing, gravitational wave standard sirens etc. ~\cite{Freedman:2000cf,Perlmutter:1998np,Riess:2016jrr,Riess:2020fzl}. The latest SH0ES measurement based on the Supernovae calibrated by Cepheids is $H_0 = 73.04 \pm 1.04 \,{\rm km}\,{\rm s}^{-1}\,{\rm Mpc}^{-1}$ at 68\% CL \cite{Riess:2021jrx}. Further improvement comes from a SH0ES measurement of distance ladder callibrated by parallaxes of Cepheids in open clusters, which combined with all anchors, yields $H_0=73.01\pm 0.99 \,{\rm km}\,{\rm s}^{-1}\,{\rm Mpc}^{-1}$ \cite{Riess:2022mme}.} 

Another type of measurement is provided by the Planck collaboration which uses temperature and polarization anisotropies in the Cosmic Microwave Background (CMB) to obtain $H_0 = 67.27 \pm 0.6\,{\rm km}\,{\rm s}^{-1}\,{\rm Mpc}^{-1}$. The discrepancy between local model-independent measurements of $H_0$ and the early-Universe CMB values  {can reach $5.3 \sigma$ and it} is one of the fundamental problems in cosmology~\cite{Schoneberg:2019wmt,DiValentino:2017gzb,DiValentino:2020zio,DiValentino:2021izs,Perivolaropoulos:2021jda,Lucca:2021dxo,Verde:2019ivm,Knox:2019rjx,Jedamzik:2020zmd,Shah:2021onj,Abdalla:2022yfr}. 

 {The Baryon Acoustic Oscillations (BAO) are sound waves in the baryon-photon plasma comprising the visible matter in the post-inflationary Universe, which froze at recombination epoch. Today, they are observed in the clustering of large-scale structures by numerous galactic surveys (SDSS, DES, WiggleZ, BOSS, etc). Due to rather simple physics of the plasma waves, the BAO can be considered as a standard ruler evolving with the Universe, thus providing another window into studying  cosmological models} ~\cite{Dunkley:2010ge,Addison:2013haa,Aubourg:2014yra,Cuesta:2014asa,Ade:2013zuv,Ade:2013gez,Story:2014hni,Ade:2015xua,Alam:2016hwk,Troxel:2017xyo,Planck:2018vyg,Cuceu:2019for,Dainotti:2021pqg}. A scale important for BAO measurements is set by the sound horizon at drag epoch. As it is known, at recombination, the photons decouple from the baryons first, at $z_* \approx 1090$, which gives rise to the CMB. The baryons stop feeling the drag of photons at the drag epoch, $z_d \approx 1059$, which sets the standard ruler for the BAO. The Planck Collaboration value of the sound horizon is $r_d^\mathrm{Pl18} = 147.09\pm0.26\,\mathrm{Mpc}$ \cite{Planck:2018vyg}, and the late-time estimation  {for it} is $r_d^\mathrm{H0LiCOW+SN+BAO+SH0ES} = 136.1\pm2.7\,\mathrm{Mpc}$ \cite{Arendse:2019hev}. Other estimations give numbers in this range, depending on the datasets in use, for example see \cite{Verde:2016ccp, Aghanim:2018eyx,eBOSS:2020yzd,Nunes:2020uex,Nunes:2020hzy}.

Many papers discuss the relation between the Hubble constant $H_0$ and the sound horizon scale $r_d$ for different models \cite{Aylor:2018drw,Knox:2019rjx, Pogosian:2020ded,Aizpuru:2021vhd}. Some claim that resolving the $H_0$ tension is not enough, since one has to also take into account the model's effect on the sound horizon. This means that  {one should rule out models that resolve the $H_0$ tension without resolving the $r_d$ tension simultaneously} \cite{Jedamzik:2020zmd,Aizpuru:2021vhd,delaMacorra:2021hoh}. Since $H_0$ and $r_d$ are strongly connected, it seems hard to disentangle them without making any assumptions.  {In order to have an independent crosscheck on DE models constraints, we remove the dependence on $H_0 \cdot r_d$ by marginalizing over it using a $\chi^2$ redefinition.} Such approach has already been used to different extent in the literature. In \cite{Lazkoz:2005sp} it has been performed on SnIa Gold dataset to compare different parametrizations of H(z). Ref \cite{Basilakos:2016nyg} study the growth index by comparing $\Lambda$CDM to several dark energy models by marginalizing over $M_B$ and $\sigma_8$. Ref \cite{Anagnostopoulos:2017iao} study different cosmological models by marginalizing over $H_0$ and find that one cannot rule out non-flat models or dynamical dark energy. They observe that the time-varying equation of state parameter $w(z)$ cannot be constrained by the current expansion data. Finally \cite{Camarena:2021jlr} use marginalization over $H_0$ and $M_B$ in different datasets to show  that a hockey-stick dark energy cannot solve the $H_0$ tension.

One possibility to resolve the tension is by changing the DE model. The question whether the DE is a constant energy density or with a dynamical behavior has been studied in different works \cite{Benisty:2021wxi,Capozziello:2011et,Bull:2015stt,DiValentino:2021izs,Yang:2021flj}. This motivates a host of DE parametrizations \cite{Wang:2018fng,Reyes:2021owe,Colgain:2021pmf,2108.04188}  to be used  in the search for deviations from  the cosmological constant, $\Lambda$, in observational data.  A justification for this can be found in numerous papers claiming that DE may resolve the Hubble tension, particularly for the Early Dark Energy models \cite{Gogoi:2020qif,Poulin:2018cxd,Sakstein:2019fmf,Tian:2021omz,Nojiri:2021dze,Seto:2021xua,Hill:2021yec}. 

In this work, we use two types of datasets for BAO and we combine them with SnIA. Then we marginalize over $H_0\cdot r_d$ and $H_0$ and $M_B$, respectively. This allows us to remove the need of taking priors on these quantities and thus it removes some of the implied assumptions on the models. By this method, we study $\Lambda$CDM, wCDM, the CPL parametrization of wwaCDM and also two emergent dark energy models: pEDE and gEDE. We show that even with this more extensive marginalization, one can see differences in the predictions of the different models  {inferred from the different datasets}. The second is particularly interesting in view of the growing sensitivity towards the implied assumptions in processing the data. We then perform a statistical analysis on the so obtained results using 4 well-established measures. We confirm that constraining $w_a$ seems impossible from this method, while the errors on $w_0$ improve  {significantly} when we add $SnIA$. Surprisingly, the different BAO datasets show different preference for the flatness of the universe.

The plan of the work is as follows: Section \ref{sec:theory} formulates the relevant theory.  Section \ref{sec:method} describes the method. Section \ref{sec:Res} shows the results with a model comparison. Finally, section \ref{sec:sum} summarizes the results.

\section{Theory}
\label{sec:theory}
A Friedmann - Lema\^itre - Robertson - Walker metric with the scale parameter $a = 1/(1+z)$ is considered, where $z$ is the redshift. The evolution of the Universe for it is governed by the Friedmann equation which connects the equation of the state for $\Lambda$CDM background: 
\begin{equation}
    E(z)^2 =  \Omega_{m} (1+z)^3 + \Omega_{K} (1+z)^2 + \Omega_{\Lambda}(z),
    \label{eq:hzlcdm}
\end{equation}
with the expansion of the Universe $E(z)^2= H(z)/H_0$, where $H(z) := \dot{a}/a$ is the Hubble parameter at redshift $z$ and $H_0$ is the Hubble parameter today. $\Omega_{m}$, $\Omega_{\Lambda}$ and $\Omega_{K}$ are the fractional densities of matter, DE and the spatial curvature at redshift $z=0$. We ignore radiation, since we take a look on the late Universe. The spatial curvature is expected to be zero for a flat Universe, $\Omega_K=0$. We can expand this simple model by considering a DE component depending on $z$. This can be done with a generalization of the Chevallier-Polarski-Linder (CPL) parametrization \cite{Chevallier:2000qy,Linder:2002et,Linder:2005ne,Barger:2005sb} of the $wwaCDM$ model:
\begin{equation}
\Omega_{\Lambda} \left(z\right) = \Omega_{\Lambda}^{(0)}  \exp\left[\int_0^{z} \frac{3(1+w(z')) dz'}{1+z'}\right]
\label{eq:integOmeLambda}
\end{equation}
in which we consider three possible models:
\begin{equation}
w(z) =\begin{cases} w_0 + w_a z & \text{Linear }  \\ 
 w_0 + w_a \frac{z}{z+1} & \text{CPL }  \\ 
w_0 - w_a \log{(z+1)} & \text{Log }  \end{cases} 
\end{equation}
which recover $\Lambda$CDM for $w_0=-1, w_a=0$.

To this parametrization we add another model, namely the phenomenologically Emergent Dark Energy (pEDE) \cite{Li:2019yem, Li:2020ybr} and its generalization (gEDE). gEDE is described by: 
\begin{equation}
\Omega_{DE}(z)=\Omega_\Lambda \frac{1-\tanh(\bar{\Delta} \log_{10}(\frac{1+z}{1+z_t}))}{1+\tanh(\bar{\Delta} \log_{10}({1+z_t})}
\end{equation}
with pEDE-CDM recovered for $\bar{\Delta}=1$, and $\Lambda CDM$ for $\bar{\Delta}=0$. The parameter $z_t$ here is the transitional redshift where ${\Omega}_{DE}(z_t)=\Omega_m (1+z_t)^3$.  Note, $z_t$ is obtained as a solution of this equation and thus, it is not a free parameter,  {but a calculated one.} The analytical form of $w(z)$ then can be obtained from the integral~(\ref{eq:integOmeLambda}), see \cite{Li:2020ybr}.

 {The BAO measurements provide different directions. The radial projection $D_H(z)= c/H(z)$ gives:}
\begin{equation}
\frac{D_H}{r_d} = \frac{c}{H_0 r_d} \frac{1}{E(z)},
\end{equation}
 {which includes the parameter $\frac{c}{H_0 r_d}$. The tangential BAO measurements are given in terms of the angular diameter distance $D_\textrm{A}$:}
\begin{align}\label{}
D_\textrm{A}
=\frac{c}{H_0}\frac{1}{(1+z)  \sqrt{|\Omega_{K}|}  } \textrm{sinn}\left[|\Omega_{K}|^{1/2}\Gamma(z)\right]\ ,
\end{align}
where $\textrm{sinn}(x) \equiv \textrm{sin}(x)$, $x$, $\textrm{sinh}(x)$ for $\Omega_{K}<0$, $\Omega_{K}=0$, $\Omega_{K}>0$ respectively. The $\Gamma$ function is defined as:
\begin{equation}
\Gamma(z) = \int \frac{dz'}{E(z')}
\end{equation}
 {where E(z) is related to the equation of state of the Universe as defined above}. Thus, the measurement $D_A/r_d$ can expressed as:
\begin{subequations}
\begin{equation}
\frac{D_A}{r_d} = \frac{c}{H_0 r_d} f(z),
\end{equation}
where:
\begin{equation}
f\left(z\right) = \frac{1}{(1+z)  \sqrt{|\Omega_{K}|}  } \textrm{sinn}\left[|\Omega_{K}|^{1/2}\Gamma(z)\right]. 
\end{equation}
\end{subequations}
 {A related quantity  {used in the radial BAO measurements} is the comoving angular diameter distance $D_M= D_A (1+z)$.}

Furthermore, we use dataset featuring the BAO angular scale measurement $\theta_{BAO}(z)$. It gives the angular diameter distance $D_A$ at the redshift z:
\begin{equation}
\theta_{BAO}\left(z\right) = \frac{r_d}{\left(1+z\right)D_A(z)} = \frac{H_0 r_d}{c}h(z),
\end{equation}
with:

 {\begin{equation}
h\left(z\right) = \frac{1}{ (1+z) f(z)}
\end{equation}}

 {We see that both $D_A/r_d$ and $\theta_{BAO}$ and $D_H/r_d$ depend on the quantity $H_0 \cdot r_d$ which can be eliminated from the corresponding $\chi^2$, as we demonstrate in the next section. }

\begin{table*}
\centering
\scalebox{1.1}{
\begin{tabular}{|c|c|c|c|c|}       
\hline\hline                                            
$z$  & $D_A/r_d$ & $\sigma_{Data}$ & year  &  Ref. \\ \hline\hline 
$0.11$   & $2.607$ & $0.138$&  $2021$  & \cite{deCarvalho:2021azj}\\
\hline 
$0.24$    & $5.594$& $0.305$&  $2016$  & \cite{BOSS:2016goe}\\
\hline 
$0.32$     &  $6.636$  &  $0.11$   &  $2016$   & \cite{BOSS:2016wmc} \\
\hline 
$0.38$     &  $7.389$  &  $0.122$  &  $2019$  &  \cite{BOSS:2016hvq}\\
\hline 
$0.44$     &  $8.19$  &  $0.77 $  &   $2012$   & \cite{Blake:2012pj}\\
\hline 
$0.51$    &  $7.893$  &  $0.279$  &   $2015$ & \cite{Carvalho:2015ica}	\\
\hline 
$0.54$     &  $9.212$  &  $0.41$    &  $2012$  & \cite{ Seo:2012xy}\\
\hline 
$0.6$     &  $9.37$  &  $0.65$   &  $2012$  & \cite{Blake:2012pj}\\
\hline 
$0.697$     &  $10.18$  &  $0.52$   &  $2020$  & \cite{Sridhar:2020czy,Gil-Marin:2020bct}*\\
\hline 
$0.73$     &  $10.42$  &  $0.73$  &   $2012$   & \cite{Blake:2012pj}\\
\hline 
$0.81$     &  $10.75$  &  $0.43$   &  $2017$   & \cite{DES:2017rfo}\\
\hline 
$0.85$     &  $10.76$  &  $  0.54$   &  $ 2020$  & \cite{Tamone:2020qrl}\\
\hline 
$0.874$     &  $11.41$  &  $0.74$    &  $2020$ & \cite{Sridhar:2020czy}\\
\hline 
$1.00$     &  $11.521$  &  $1.032$    &  $2019$ & \cite{Zhu:2018edv}\\
\hline 
$1.480$     &  $12.18$  &  $0.32$  &  $2020$  & \cite{Hou:2020rse,Gil-Marin:2020bct,Bautista:2020ahg}* \\
\hline 
$2.00$     &  $12.011$  &  $0.562$  &   $2019$  & \cite{Zhu:2018edv}\\
\hline 
$2.35$     &  $10.83$  &  $0.54$  &  $2019$   & \cite{Blomqvist:2019rah}, \cite{duMasdesBourboux:2020pck}*  \\
\hline 
$2.4$     &  $10.5$  &  $0.34$  & 2017 & \cite{duMasdesBourboux:2017mrl}\\
\hline\hline                                                
\end{tabular} 
}
\caption{{\it{A compilation of BAO measurements from  diverse releases of the SDSS, WiggleZ, DES etc.  {Values marked with * are calculated trough their covariance matrices relating $D_M$ and $D_H$} }}}
\label{tab:data1}                                           
\end{table*}

\begin{table} 
\centering
\scalebox{1.1}{
\begin{tabular}{|c|c|c|c|c|}       
\hline\hline     
z & $\theta$ & $\sigma_\theta$  & Ref. \\
			\hline
			0.11 & 19.80 & 3.26  & \cite{deCarvalho:2020ftb}\\
			\hline
			0.235 & 9.06 & 0.23  & \cite{Alcaniz:2016ryy}\\
			\hline
			0.365 & 6.33 & 0.22 & \cite{Alcaniz:2016ryy} \\
			\hline
			0.450 & 4.77 & 0.17 & \cite{Carvalho:2015ica} \\
			\hline
			0.470 & 5.02 & 0.25 & \cite{Carvalho:2015ica}\\
			\hline
			0.490 & 4.99 & 0.21  & \cite{Carvalho:2015ica}\\
			\hline
			0.510 & 4.81 & 0.17  & \cite{Carvalho:2015ica}\\
			\hline
			0.530 & 4.29 & 0.30  & \cite{Carvalho:2015ica}\\
			\hline
			0.550 & 4.25 & 0.25  & \cite{Carvalho:2015ica}\\
			\hline
			0.570 & 4.59 & 0.36  & \cite{Carvalho:2017tuu} \\
			\hline
			0.590 & 4.39 & 0.330  & \cite{Carvalho:2017tuu} \\
			\hline
			0.610 & 3.85 & 0.31  & \cite{Carvalho:2017tuu} \\
			\hline
			0.630 & 3.90 & 0.43  & \cite{Carvalho:2017tuu}\\
			\hline
			0.650 & 3.55 & 0.16 & \cite{Carvalho:2017tuu}\\
			\hline
			2.225 & 1.77 & 0.31 & \cite{deCarvalho:2017xye} \\
			\hline\hline  
		\end{tabular}
}
\caption{{\it{A compilation of angular BAO measurements from luminous red and blue galaxies and quasars from diverse releases of the SDSS. The $BAO{_\theta}$ dataset were taken from \cite{Nunes:2020hzy}.}}}
\label{tab:data2}                                           
\end{table}

Finally, we add the type Ia supernovae (SnIA) measurements, described by the luminosity distance $\mu(z)$. It is related to the Hubble parameter through the angular diameter distance as  $D_A=d_L(z)/(1+z)^2$. For the SnIA standard candles, the distance modulus $\mu(z)$ is related to the luminosity distance through
\begin{equation}
	\mu_B (z) - M_B = 5 \log_{10} \left[ d_L(z)\right] + 25  \,.
\label{eq:dist_mod_def}
\end{equation}
where $d_L$ is measured in units of Mpc, and $M_B$ is the absolute magnitude. There is a degeneracy between $H_0$ and $M_B$, in such a way that total absolute magnitude reads: $M_B + 25 + 5 \log_{10} \left(\frac{c/H_0}{Mpc}\right)$. This degeneracy, can also be used to remove the dependence on $H_0$ and $M_B$ in the $\chi^2$.

\section{Method}
\label{sec:method}

 {In order to infer the parameters of certain model from the observations, one needs to define the appropriate $\chi^2$}. The goal of our analysis is to redefine the corresponding $\chi^2$ in all datasets, in a way that eliminates the dependence on degenerate parameters, such as $H_0\cdot r_d$ (or $H_0$ and $M_B$ for SnIA), but maintains the dependence on the equation of state that enter into $\Gamma(z)$. 

\subsection{BAO redefinition}
 {A DE model includes n-free parameters (i.e. $\Omega_m, \Omega_K, w_0, w_a ... $), constrained by minimizing the $\chi^2$:}
\begin{equation}
\begin{split}
\chi^2 = \sum_{i} \left[\vec{v}_{obs} - \vec{v}_{model}\right]^{T} C_{ij}^{-1} \left[\vec{v}_{obs} -  \vec{v}_{model} \right] 
\end{split}
\end{equation}
where $\vec{v}_{obs}$ is a vector of the observed points  {at each $z$ (i.e. $D_M/r_d$, $D_H/r_d$, $D_A/r_d$ or $\theta_{BAO}$)} and $\vec{v}_{model}$ is the theoretical prediction of the model. It is possible to rewrite the vector as the dimensionless function multiplied by the $\frac{c}{H_0 r_d}$ parameter:  
\begin{equation}
\vec{v}_{model} = \frac{c}{H_0 r_d}\left(f(z) , E(z)^{-1}\right) = \frac{c}{H_0 r_d} \vec{f}_{model}.
\end{equation}
$C_{ij}$ is the covariance matrix. For uncorrelated points  { the covariance matrix is a diagonal matrix, and its elements are the inverse errors $\sigma_i^{-2}$.} The statistics of the BAO is not fully a Gaussian but we consider this as an approximation. Following the approach in \cite{Lazkoz:2005sp,Basilakos:2016nyg,Anagnostopoulos:2017iao,Camarena:2021jlr},  {one can isolate $\frac{c}{H_0 r_d}$ in the $\chi^2$ by writing it as}:
\begin{equation}
\chi^2 = \left(\frac{c}{H_0 r_d}\right)^2 A - 2 B \left(\frac{c}{H_0 r_d}\right) + C,
\end{equation} 
where:
\begin{subequations}
\begin{equation}
A  = f^j(z_i) C_{ij} f^i(z_i),
\end{equation}
\begin{equation}
B = \frac{f^j(z_i) C_{ij} v_{model}^i(z_i) + v_{model}^j(z_i) C_{ij} f^i(z_i)}{2},
\end{equation}
\begin{equation}
C = v_j^{model} C_{ij} v_i^{model}.
\end{equation}
\label{eq:termChi2}
\end{subequations}
Using Bayes’s theorem and marginalizing over $c/\left(H_0 r_d\right)$, we arrive at:
\begin{equation}
p\left(D,M\right) = \frac{1}{p\left(D|M\right)} \int \exp\left[-\frac{1}{2}\chi^{2}\right] d\frac{c}{H_0 r_d},
\end{equation}
where $D$ is the data we use, and the $M$ is the model. Consequently,   {using $\tilde{\chi}^2_{BAO} = -2 \ln p\left(D,M\right)$ we get} the marginalized $\chi^2$:
\begin{equation}
\tilde{\chi}^2 = C-\frac{B^2}{A} + \log\left(\frac{A}{2 \pi}\right),
\label{eq:chi2BAO}
\end{equation} 
 {This last equation is the final $\chi^2$ we use. For it, due to the marginalization procedure, the $\chi^2$ depends only on $f(z)$ and $h(z)$ which do not include $H_0$ and $r_d$ inside. }

\subsection{$\theta_{BAO}$ data}
We use the same approach for the $\theta_{BAO}(z)$ measurements:
\begin{equation}
\chi^2_{\theta,BAO}  = \sum_{i = 1}^{N} \left( \frac{\theta(z_i) - {\theta}_{D}^{i}}{\sigma_i} \right)^2,
\end{equation}
where ${\theta}_{D}^{i}$ and $\sigma_i$ are the observational data and the corresponding uncertainties at the observed redshift $z_i$. The reconstructed $\chi^2_{\theta,BAO}$, then, is the following:
\begin{equation}
\chi^2_{\theta,BAO} = \left(\frac{H_0 r_d}{c}\right)^2 A_{\theta} - 2 B_{\theta}  \left(\frac{H_0 r_d}{c}\right) + C,
\end{equation}
where:
\begin{subequations}
\begin{equation}
A_{\theta}  = \sum_{i = 1}^{N} \frac{h(z_i)^2}{\sigma_i^2},
\end{equation}
\begin{equation}
B_{\theta}  = \sum_{i = 1}^{N} \frac{{\theta}_{D}^{i}\, h(z_i) }{\sigma_i^2},
\end{equation}
\begin{equation}
C_{\theta}  = \sum_{i = 1}^{N} \frac{\left({\theta}_{D}^{i}\right)^2}{\sigma_i^2}.
\end{equation}
\end{subequations}
Using Bayes’s theorem and marginalizing over $H_0 r_d/c$, we arrive at the marginalized $\chi^2$,  {which} is the same as in Eq~(\ref{eq:chi2BAO}), only with A, B and C now functions of $\theta$. This $\tilde{\chi}^2_{\theta}$ also depends only on $h(z)$, without any dependence on $H_0 \cdot r_d/c$.

\subsection{Supernova redefinition}
Following the approach used in \cite{DiPietro:2002cz,Nesseris:2004wj,Perivolaropoulos:2004yr,Lazkoz:2005sp} we assumed no prior constraint on $M_B$,
which is just some constant and we integrated the probabilities over $M_B$. The integrated $\chi^2$ yields:
\begin{equation}
\tilde{\chi}^2_{SN} = D-\frac{E^2}{F} + \ln\frac{F}{2\pi},
\end{equation}
where:
\begin{subequations}
\begin{equation}
D = \sum_i \left(\frac{\mu_{}^{i} - 5 \log_{10}\left[d_L(z_i)\right]}{\sigma_i}\right)^2, 
\end{equation}
\begin{equation}
E = \sum_i \frac{\mu_{}^{i} - 5 \log_{10}\left[d_L(z_i)\right]}{\sigma_i^2},
\end{equation}
\begin{equation}
F = \sum_i \frac{1}{\sigma_i^2}.
\end{equation}
\end{subequations}
Here $\mu_{}^{i}$ is the observed luminosity, $\sigma_i$ is its error and the $d_L(z)$ is the luminosity distance. The values of $M$ and $H_0$ don't change the marginalized $\tilde{\chi}^2_{SN}$.  {In order to use the covariance matrix provided for the Pantheon dataset one needs to transform $D,E,F$ as follows:}
\begin{subequations}
\begin{equation}
D = \sum_i \left( \Delta\mu \, C^{-1}_{cov} \, \Delta\mu^T \right)^2,
\end{equation}
\begin{equation}
E = \sum_i \left( \Delta\mu \, C^{-1}_{cov} \, E \right),
\end{equation}
\begin{equation}
F = \sum_i  C^{-1}_{cov}  .
\end{equation}
\end{subequations}
where $\Delta\mu =\mu_{}^{i} - 5 \log_{10}\left[d_L(z_i)\right)$, $E$ is the unit matrix and $C^{-1}_{cov}$ is the inverse covariance matrix of the dataset. The total covariance matrix is given by $C_{cov}=D_{stat}+C_{sys}$, where $D_{stat}=\sigma_i^2$ comes from the measurement and $C_{sys}$ is provided separately \cite{Deng:2018jrp}. Notice that the form of $\tilde{\chi}^2_{BAO}$ and $\tilde{\chi}^2_{SN}$ is a bit different, since for the $\tilde{\chi}^2_{BAO}$ we remove the dependence of $c/{H_0 r_d}$ which multiply the $f(z)$ and in the case of $\tilde{\chi}^2_{SN}$ the parameter $\bar{M}$ is added the total value of $\mu$.

In our analysis we also consider the combined likelihood 
\begin{equation}
\tilde{\chi}^{2} = \tilde{\chi}_{BAO}^{2} + \tilde{\chi}_{SN}^{2}.
\end{equation}
Here $\tilde{\chi}_{BAO}^{2}$ stands for the $BAO$ or for the $BAO_{\theta}$ datasets independently. The distinction between the hyper-parameters quantifying uncertainties in a dataset and the free parameters of the cosmological model is purely conceptual.  {It is important to note that the so defined $\chi^2$ is not normalized  {because of which} its absolute value is not a useful measure of the quality of a given fit. Moreover, it is biased towards larger number of parameters and not very good for small datasets, such as the ones we use \cite{Lazkoz:2005sp}. For this reason, we use it only to calculate the more balanced statistical measures, see below.  }

\subsection{Datasets and priors}
In this work, we consider two different BAO datasets, to which we add the binned Panthon supernovae dataset with its covariance matrix. The BAO datasets can be found summarized in Table~(\ref{tab:data1}) and Table~(\ref{tab:data2}) .

 {The first BAO dataset, shown on Table \ref{tab:data1} and denoted $BAO$,  {contains a combination of various angular measurements, to which we add points from the most recent to date eBOSS data release (DR16), which come as  angular ($D_M$) and radial ($D_H$) measurements and their covariance}. The points and the covariance matrices can be found in \cite{Cao:2022ugh}. This choice of points allows us to integrate the quantity $H_0 \cdot r_d$ by summing the corresponding $\chi^2$ of the two types of measurements. While the covariance for some points is known and we include it, for the rest, we have to additionally test for possible correlations. To do so we use the approach from \cite{Kazantzidis:2018rnb}, which we also used in \cite{Benisty:2020otr}.} It consists of adding random correlation terms in the covariance matrix and testing the effect on the final result. Explicitly, we use $$\sigma_{ii}\to \sigma_{ii}+ \sigma_i \sigma_j/2,$$ where $\sigma_i$ is the $1\sigma$ error of the points.  Applying the procedure shows that the points can be considered  "effectively uncorrelated" which allows us to use them to infer the cosmological parameters. Even if there are small correlations, the procedure shows the small correlations don't affect the final result considerately.

The second dataset shown on Table \ref{tab:data2}, denoted $BAO{_\theta}$, consists of 15 points, coming from transversal BAO measurements \cite{Nunes:2020hzy}. Importantly, the transversal BAO analysis does not need to assume a fiducial cosmology, particularly on the $\Omega_K$ parameter which is included in the standard BAO analysis \cite{Nunes:2020hzy}. These points are claimed to be uncorrelated, however, using this cosmology-independent methodology means that their errors are larger than the errors obtained using the standard fiducial cosmology approach. One should note that using a fiducial cosmology is accounted for by the Alcock-Paczynski distortion \cite{Lepori:2016rzi}, so it does not compromise the integrity of the first dataset. However, we would like to investigate the over-all effect of intrinsic assumptions in the final results and to check if the two datasets are equivalent in this respect.

Finally, we add the Pantheon dataset which contains $1048$ supernovae luminosity measurements in the redshift range $z\in (0.01,2.3)$ \cite{Pan-STARRS1:2017jku} binned into 40 points. To the statistical error we add also the systematic errors as provided by the binned covariance matrix \footnote{\url{https://github.com/dscolnic/Pantheon/}}.

We perform the $H_0\cdot r_d$-integration procedure, outlined in previous sections, first on the two different BAO datasets alone, and then on the combination of the appropriate BAO dataset plus the Pantheon dataset. The priors we use are: $\Omega_m \in (0.2,0.4)$, $w_0\in  (-2,-0)$, $w_a \in(-2,1)$, $\Omega_K \in (-0.3,0.3)$. We set $\Omega_{\Lambda}^{(0)} = 1- \Omega_m -\Omega_K$. For gEDE we use the redefinition $\Delta=-\bar{\Delta}, w_a=z_t$, so that it can be plotted on the same plots as the other models. As mentioned before $z_t$ is not a free parameter  {thus it is not a parameter in the MCMC} and it is found by solving the appropriate transcendental equation using the package {\textit{sympy}}. Regarding the problem of likelihood maximization, we use an affine-invariant Markov Chain Monte Carlo (MCMC) nested sampler, as it is implemented within the open-source package {\textit{Polychord}} \cite{Handley:2015fda} with the {\textit{GetDist}} package \cite{Lewis:2019xzd} to present the results.  {In \textit{Polychord} convergence is defined as when the posterior mass contained in the live points is $p=10^{-2}$ of the total calculated evidence. We check that our chains are stable with respect to changes in the parameter $p$ and furthermore by checking the Geweke score and the Gelmen-Rubin diagnostic with the package {\textit{pymcmcstat}}.  }

\begin{table*}
	\begin{center}

	\begin{tabular}{|c|c|c|c|c|c|c|c|c|}
			\hline
		Model & $\Omega_m$ & $\Omega_K$ & $w_0$ & $w_a$ & $\Delta$AIC & $\Delta BIC$ & $\Delta$DIC & ln(BF)  \\
			\hline \hline
			$\mathbf{BAO}$ & & & & & & & & \\
			\hline
			LCDM & $0.314\pm 0.014$ & - & - & - & 0 & 0 & 0 & 0 \\
			\hline
			wCDM & $0.292\pm 0.027$ & - & $-0.658\pm 0.119$ & - & -1.184 & -2.228 & 0.810 & 1.693 \\
			\hline
			wwaCDM & $0.314\pm 0.053$ & - & $-0.644\pm 0.135$ & $-0.181\pm 0.3$ & -3.062 & -5.151 & 0.869 & 0.375 \\
			\hline
			OkLCDM & $0.321\pm 0.015$ & $-0.061\pm 0.053$ & - & - & -3.007 & -4.052 & -1.216 & -0.835 \\
			\hline
			Linear & $0.315\pm 0.051$ & - & $-0.63\pm 0.127$ & $-0.196\pm 0.293$ & -2.995 & -5.084 & 0.930 & 0.241 \\
			\hline
			CPL & $0.293\pm 0.054$ & - & $-0.662\pm 0.173$ & $-0.061\pm 0.606$ & -3.299 & -5.388 & 0.600 & 1.267 \\
			\hline
			Log & $0.308\pm 0.046$ & - & $-0.651\pm 0.149$ & $0.153\pm 0.376$ & -3.142 & -5.231 & 0.786 & 0.639 \\
			\hline
			pEDE & $0.31\pm 0.015$ & - & - & - & 0.717 & 0.717 & 0.486 & -3.796 \\
			\hline
			gEDE & $0.311\pm 0.016$ & - & $-0.278\pm 0.209$ & 0.290 & -1.728 & -2.773 & 0.161 & -1.900 \\
			\hline \hline
			$\mathbf{BAO_{\theta}}$ & & & & & & & & \\
			\hline
	LCDM & $0.325\pm 0.057$ & - & - & - & 0 & 0 & 0 & 0 \\
			\hline 
			wCDM & $0.324\pm 0.064$ & - & $-0.929\pm 0.356$ & - & -1.837 & -2.545 & 0.113 & -0.545 \\
			\hline
			wwaCDM & $0.319\pm 0.058$ & - & $-0.89\pm 0.43$ & $-0.314\pm 0.73$ & -3.916 & -5.332 & 0.067 & -0.538 \\
			\hline
			OkLCDM & $0.327\pm 0.053$ & $0.038\pm 0.181$ & - & - & -2.075 & -2.783 & -0.084 & 0.092 \\
			\hline
			Linear & $0.324\pm 0.064$ & - & $-0.872\pm 0.449$ & $-0.337\pm 0.963$ & -3.821 & -5.237 & 0.118 & -0.467 \\
			\hline
			CPL & $0.327\pm 0.063$ & - & $-0.854\pm 0.438$ & $-0.448\pm 0.932$ & -3.791 & -5.207 & 0.141 & -0.628 \\
			\hline
			Log & $0.316\pm 0.066$ & - & $-1.07\pm 0.432$ & $-0.527\pm 0.97$ & -3.853 & -5.269 & 0.094 & -0.597 \\
			\hline
			pEDE & $0.341\pm 0.051$ & - & - & - & 0.165 & 0.165 & 0.114 & -0.332 \\
			\hline
			gEDE & $0.343\pm 0.044$ & - & $-0.877\pm 0.693$ & 0.214 & -1.916 & -2.624 & 0.060 & -0.184 \\
			\hline
		\end{tabular}
	\end{center}
	\caption{Constraints at 68\% CL errors on the cosmological parameters for the different tested models for the two BAO only datasets: $BAO$ and $BAO{_\theta}$}
\label{tab:res1}   
\end{table*}

\begin{table*}
	\begin{center}
		\begin{tabular}{|c|c|c|c|c|c|c|c|c|}
			\hline
			Model & $\Omega_m$ & $\Omega_K$ & $w_0$ & $w_a$ & $\Delta$AIC & $\Delta BIC$ & $\Delta$DIC & ln(BF) \\
			\hline
			$\mathbf{BAO+SN}$ & & & & & & & &\\
			\hline
			LCDM & $0.305\pm 0.011$ & - & - & - & 0 & 0 & 0 & 0 \\
			\hline
			wCDM & $0.302\pm 0.012$ & - & $-0.986\pm 0.045$ & - & -1.603 & -3.714 & 0.240 & -2.777 \\
			\hline
			wwaCDM & $0.361\pm 0.034$ & - & $-1.18\pm 0.139$ & $-0.376\pm 0.672$ & -22.4 & -26.6 & -18.6 & 16.9 \\
			\hline
			OkLCDM & $0.336\pm 0.018$ & $-0.211\pm 0.066$ & - & - & -20.9 & -23.1 & -19.1 & 18.5 \\
			\hline
			Linear & $0.333\pm 0.084$ & - & $-1.128\pm 0.118$ & $-0.125\pm 1.056$ & -22.4 & -26.7 & -18.6 & 16.6 \\
			\hline
			CPL & $0.369\pm 0.021$ & - & $-1.166\pm 0.134$ & $-0.569\pm 0.889$ & -22.2 & -26.4 & -18.4 & 16.4 \\
			\hline
			Log & $0.335\pm 0.055$ & - & $-1.183\pm 0.111$ & $-0.2\pm 0.865$ & -22.0 & -26.2 & -18.2 & 16.3 \\
			\hline
			pEDE & $0.353\pm 0.014$ & - & - & - & -18.1 & -18.1 & -18.2 & 18.4 \\
			\hline
			gEDE & $0.36\pm 0.022$ & - & $-1.319\pm 0.478$ & 0.176 & -20.3 & -22.4 & -18.4 & 18.3 \\
			\hline
		\hline
			$\mathbf{BAO_{\theta}+SN}$ & & & & & & & & \\
			\hline
		LCDM & $0.3\pm 0.014$ & - & - & - & 0 & 0 & 0 & 0 \\
			\hline
			wCDM & $0.321\pm 0.049$ & - & $-1.084\pm 0.141$ & - & -1.765 & -3.772 & 0.144 & -1.650 \\
			\hline
			wwaCDM & $0.338\pm 0.044$ & - & $-1.095\pm 0.091$ & $-0.311\pm 0.739$ & -3.978 & -7.992 & -0.068 & -1.742 \\
			\hline
			OkLCDM & $0.312\pm 0.027$ & $-0.089\pm 0.147$ & - & - & -1.918 & -3.926 & 0.014 & -0.229 \\
			\hline
			Linear & $0.33\pm 0.061$ & - & $-1.072\pm 0.117$ & $-0.279\pm 0.817$ & -3.739 & -7.753 & 0.119 & -2.107 \\
			\hline
			CPL & $0.332\pm 0.052$ & - & $-1.071\pm 0.124$ & $-0.344\pm 0.922$ & -3.707 & -7.721 & 0.213 & -1.901 \\
			\hline
			Log & $0.318\pm 0.062$ & - & $-1.087\pm 0.126$ & $-0.027\pm 0.653$ & -3.794 & -7.808 & 0.052 & -1.763 \\
			\hline
			pEDE & $0.348\pm 0.014$ & - & - & - & -0.037 & -0.037 & -0.061 & -0.120 \\
			\hline
			gEDE & $0.339\pm 0.027$ & - & $-0.892\pm 0.64$ & 0.219 & -2.047 & -4.055 & 0.015 & 0.062 \\
			\hline
		\end{tabular}
	\end{center}
	\caption{Constraints at $68\%$ CL errors on the cosmological parameters for the different tested models for the two BAO + SN  datasets: $BAO+SN$ and $BAO{_\theta}+SN$}
	\label{tab:res2}
\end{table*}

\section{Results}
\label{sec:Res}
\subsection{Posterior Distributions}
Figures \ref{fig:bwwa_BAO1},\ref{fig:bwwa_BAO2},\ref{fig:bwwa_BAO3},\ref{fig:bwwa_BAO4},\ref{fig:bwwa_BAO5} and in the Appendix show the final values obtained by running MCMC on the selected priors for the two different datasets, with the numerical values in the tables II-VI. Since we integrate $H_0$ and $r_d$, the only physically measured parameter which remains is $\Omega_m$. We see that in all the cases $\Omega_m$ is rather well constrained, even from the BAO-only datasets. The $BAO_\theta$ as expected gives larger errors which the inclusion of supernova data improves. The closest to the Planck measurement of $\Omega_m=0.315 \pm 0.007$ \cite{Aghanim:2018eyx} is the Log model for $BAO_\theta$ and the LCDM model for $BAO$, and $\Lambda$CDM/OkCDM for $BAO +SN$  and  $BAO_{\theta} +SN$ with the Log model being very close for the latter.

When we consider the other parameters, we see that the BAO only datasets are not able to limit them properly. While the $BAO_\theta$ dataset values contain $w_0=-1$ within 1 $\sigma$, the values for $BAO$ infer $w_0>-1$. Adding the SN datasets improves the constraints significantly. With respect to the parameter $w_a$ the inferred values have very big errors. When it comes to $\Omega_K$, $BAO_{\theta}$ gives values closer to a flat universe, while $BAO$ points to $\Omega_K<0$ (a closed universe). 

The two emergent dark energy models perform well in all the cases. pEDE has an error similar to $\Lambda$CDM, but at higher $\Omega_m$. gEDE also prefers higher values for $\Omega_m$. 
 
 As mentioned in the Theory part, $\Delta=0$ recovers $\Lambda$CDM, while $\Delta=-1$ recovers pEDE. We see from Fig. 4, that $\Lambda$CDM is preferred only by $BAO$, while the other datasets prefer pEDE (i.e. $\Delta$ closer to $-1$) but with large error. On the other hand, $z_t$ is consistent with the known results for $z_t\sim 0.2$. Note that in the tables and in the Appendix, we denote $\Delta \to w_0$ and $z_t \to w_a$ for notation consistency with the other models.

The conclusion from our results is that the BAO-alone datasets are useful mostly for constraining $\Omega_m$ and to lesser extent $w_0$, while they are much less sensitive to the other parameters - $w_a$ or $\Omega_k$. The BAO + SN datasets seem to give much better constraints on the DE parameters. Also, one can see that the $BAO_\theta$ dataset includes the $\Omega_k$ value of a flat universe, while the $BAO$ dataset seems to exclude it at 68\% CL.

From the the Gaussians we see that some DE models have multiple peaks, speaking of some degeneracy. The results do not seem to change with increasing the number of live points, hinting that this is a property of the models themselves or of the selected datasets. 

\begin{figure*}[ht]
\begin{tabularx}{\textwidth}{|Y|Y|}
\hline
\includegraphics[width=0.47\textwidth]{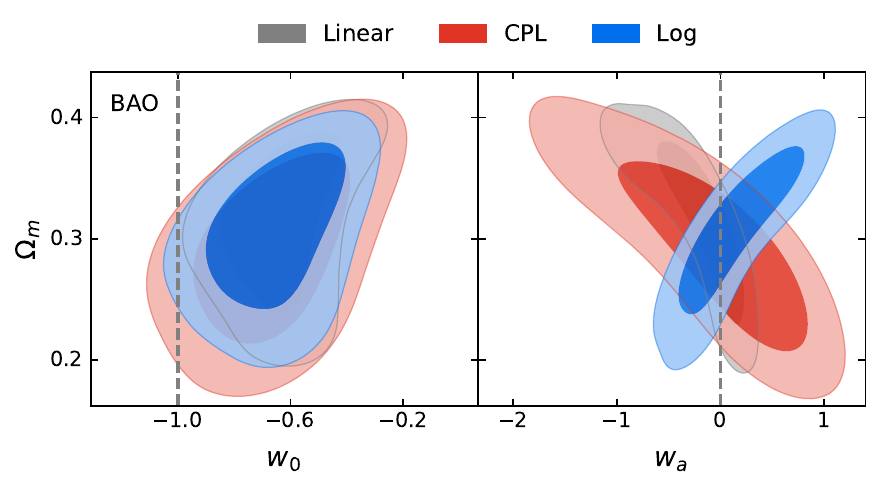}&
\includegraphics[width=0.47\textwidth]{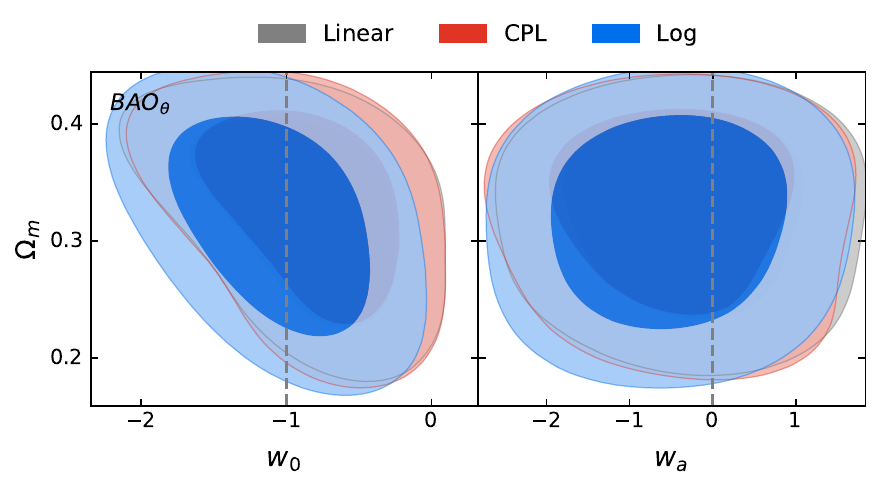}
\\
\includegraphics[width=0.47\textwidth]{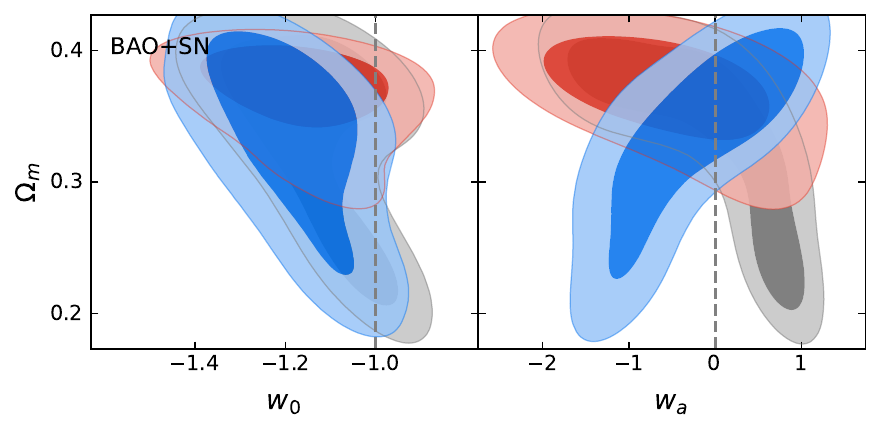}&
\includegraphics[width=0.47\textwidth]{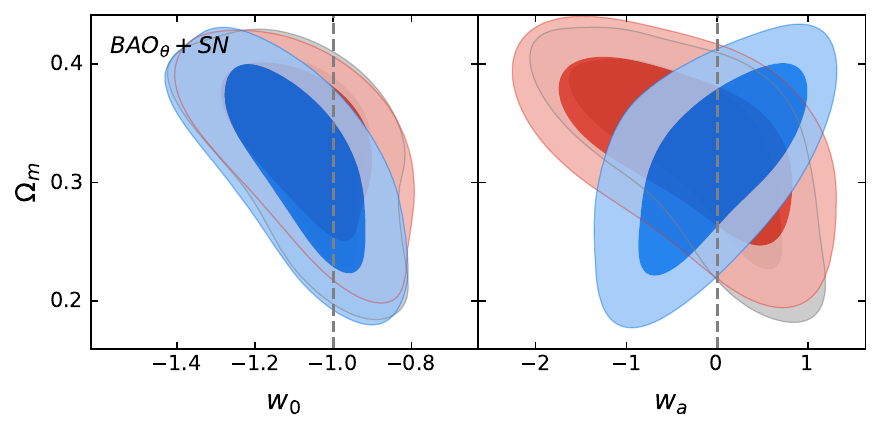}\\
\hline
\end{tabularx}
\caption{\it{The posterior distribution for $\Omega_m$ and $w_0, w_a$ for different parametrizations of the $wwaCDM$ model with the $BAO$ and $BAO_\theta$ datasets to the left and to the right, and with the Pantheon data added to the bottom panel}}
\label{fig:bwwa_BAO1}
\end{figure*}

\begin{figure}
 	\centering
\includegraphics[width=0.45\textwidth]{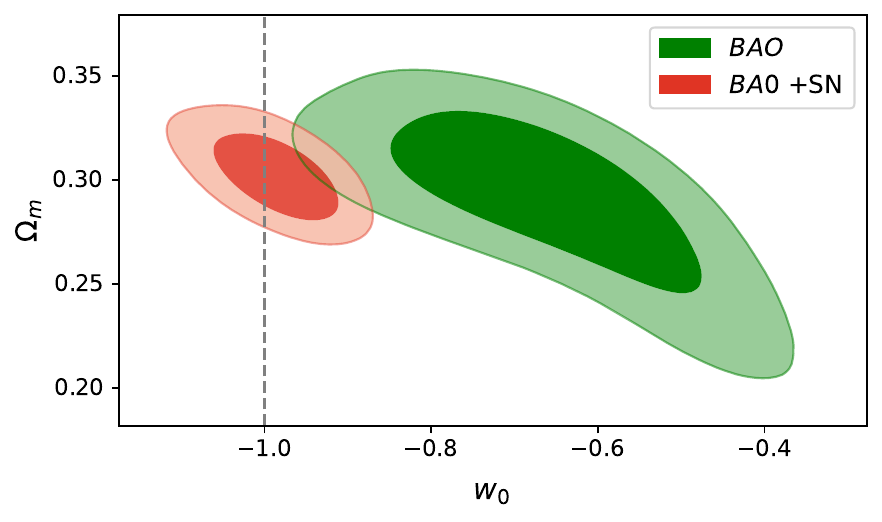}
\includegraphics[width=0.45\textwidth]{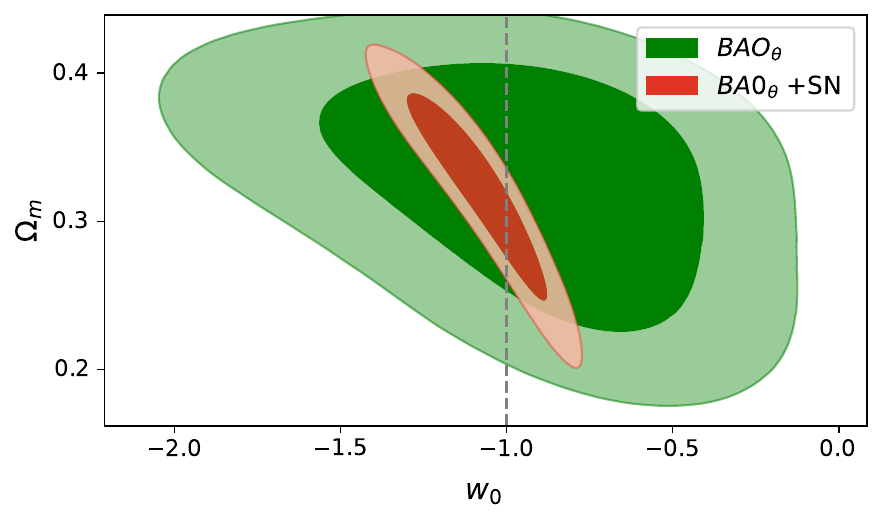}
\caption{\it{The posterior distribution for $\Omega_m$ and $w_0$ for the $wCDM$ model with the $BAO$  data on the upper and the $BAO_\theta$ data to the lower}}
 	\label{fig:bwwa_BAO2}
\end{figure}

\begin{figure}
 	\centering
\includegraphics[width=0.45\textwidth]{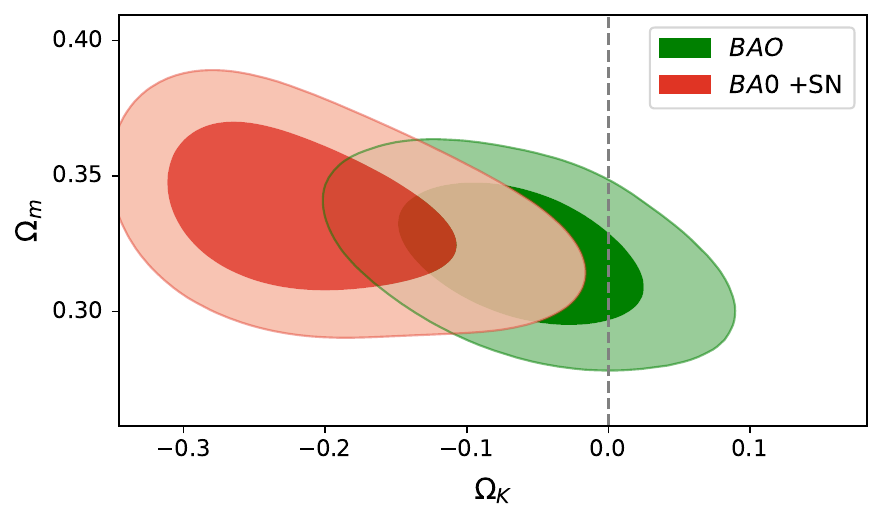}
\includegraphics[width=0.45\textwidth]{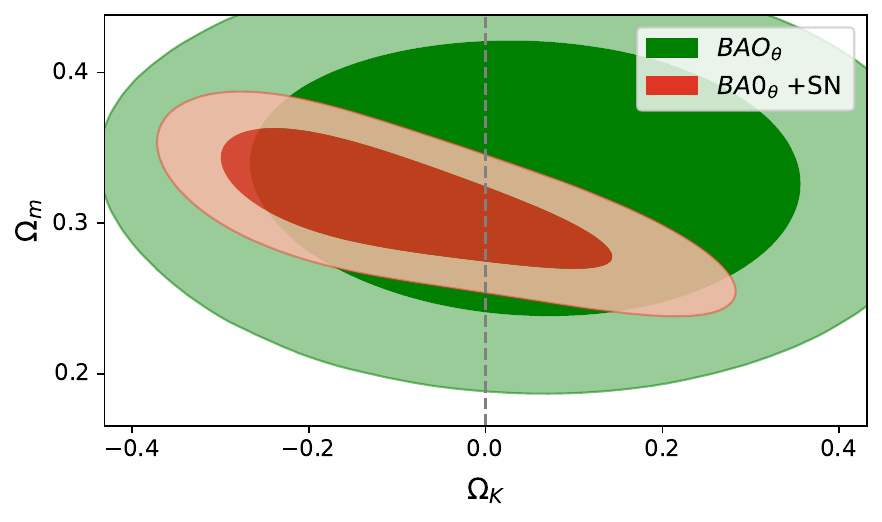}
\caption{\it{The posterior distribution for $\Omega_m$ and $\Omega_k$ for $\Omega_K$LCDM model with the $BAO$  data on the upper and the $BAO_\theta$ data to the lower}}
 	\label{fig:bwwa_BAO3}
\end{figure}

\subsection{Model Selection}
To compare the different models, we use different well-known statistical measures. We use the Akaike Information Criterion (AIC), and Bayesian Information Criterion (BIC), the Deviance Information 
Criterion (DIC) and the Bayes Factor (BF) \cite{Liddle:2007fy}. 

The AIC criterion is defined as:
        \begin{equation}
        \text{AIC}=-2\ln(\mathcal{L}_{\text{max}})+2k+
        \frac{2k(k+1)}{N_{\rm tot}-k-1}\,,
        \end{equation}
where $\mathcal{L}_{\text{max}}$ is the maximum likelihood of the data  under consideration and $N_{\rm tot}$ is the total number of data points and $k$ is the number of parameters. For large $N_{\rm tot}$, this expression
reduces to $\text{AIC}\simeq-2\ln(\mathcal{L}_{\text{max}})+2k$, which is the standard form of the AIC criterion \cite{Liddle:2007fy}. 
        
The BIC criterion is an estimator of the Bayesian evidence, (e. g \cite{Liddle:2007fy}), and is given as
\begin{equation}
\text{BIC} = -2\ln(\mathcal{L}_{\text{max}})+k \,{\rm log}(N_{\text{tot}})\,.
\end{equation}
The AIC and BIC criteria employ only the likelihood value at maximum. Since we evaluate this $\mathcal{L}_{max}$ numerically, from the Bayesian analysis, one needs to use sufficiently long chains to ensure the accuracy of $\mathcal{L}_{max}$ when evaluating AIC and BIC. The Deviance Information Criterion (DIC) \cite{Liddle:2007fy} provides all the information obtained from the likelihood calls during the maximization procedure. The DIC estimator is defined as,
\begin{equation}
{\rm DIC} =  2\overline{(D(\theta))}-D(\overline{\theta}) 
\end{equation}
where $\theta$ is the vector of parameters being varied in the model, the overline denotes the usual mean value and $D(\theta) = -2\ln(\mathcal{L(\theta)})+C$, where $C$ is a constant. We use these definitions to form the difference in the IC values of the default model ($\Lambda$CDM) and the other suggested models. I.e. we calculate $\Delta \text{IC}_{\text{model}}=\text{IC}_{\text{$\Lambda$CDM}}-\text{IC}_{\text{model}}$. The model with the minimal AIC is considered best,  \cite{Jeffreys:1939xee}, so a positive $\Delta$IC will point to a preference towards the DE model, negative -- towards $\Lambda$CDM with $|\Delta\text{IC}|\geq 2$ signifying a possible tension, $|\Delta\text{IC}|\geq 6$ -- a medium tension, $\Delta\text{IC}\geq 10$ -- a strong tension. Finally we use the Bayes factor, defined as:
$$B_{ij}= \frac{p(d|M_i)}{p(d|M_j)}$$
where $p(d|M_i)$ is the Bayesian evidence for model $M_i$. The evidence is difficult to calculate analytically, but in polychord, it is calculated numerically by the algorithm. In the tables below, we use the $ln(B_{0i})$ where "0" is $\Lambda$CDM, which we compare with all the other models (denoted by the index "i"). According to the Jeffry's scale \cite{Jeffreys:1939xee}, $ln(B_{ij}) <1$ is inconclusive for any of the models, 1-2.5 gives weak support for the model "i", 2.5 to 5 is moderate and $>5$ is strong evidence for the model "i". A minus sign gives the same for model "j".

The so defined statistical measures for the two datasets are presented in tables \ref{tab:res1} and \ref{tab:res2}. In summary, the model comparison for the different datasets gives:
\begin{itemize}
    \item For the $BAO$ dataset:  the best model from AIC, BIC and DIC is $\Lambda$CDM, followed closely (within $<1$ IC units) by pEDE. The BF agrees on that, with pEDE and gEDE being close to it. OkLCDM is comparable to LCDM.  
    
    \item For the $BAO$ + SN dataset:  the best model is $\Lambda$CDM from all IC measures. BF agrees with that for most models, with inconclusive preference for wCDM ($ln(BF)<-1$).   
   
    \item For the $BAO{_\theta}$ dataset, the best model for AIC and BIC is pEDE followed by $\Lambda$CDM. For DIC the best model is CPL, with all wCDM and wwaCDM models being better than $\Lambda$CDM. The IC difference, however, is too small to signify any tension. The BF agrees with DIC, with CPL model being best, $\Lambda$CDM - the worst. Again, inconclusively. 
    
    \item  For the $BAO{_\theta}$ + SN dataset, with respect to the AIC and BIC, the best model is $\Lambda$CDM, but pEDE is very close to it. With respect to DIC, all the models give better results than $\Lambda$CDM, with CPL - best,  but the statistical significance is extremely low. With respect to BF, however, the 3 parametrizations of wwaCDM give best results, with values representing a weak but non-negligible support. 
\end{itemize}

From this comparison we see that first, the use of statistical measures does not give entirely consistent view on the selecting the best model. This can be due to a number of factors - slow convergence of some of the models, priors not having the similar weight etc. 

Second, the two BAO datasets have preferences for different models. This may be due to different intrinsic assumptions with which the measurements have been made. The $BAO{_\theta}$ dataset, despite the larger errors, seems to give consistent results, with some weak support for DE models in the different measures. The more standard AIC and BIC, however, are always in favor of $\Lambda$CDM, with pEDE being close behind. The $BAO$ dataset seems to always prefer $\Lambda$CDM in most measures.  

We can conclude that from the two datasets of BAO points, only the $BAO$ dataset has a strong preference for $\Lambda$CDM.  Adding the Pantheon dataset to it boosts this preference to statistical significance. 
The fact that $\Lambda$CDM is not the best model statistically in all of the cases for the BAO-only datasets, may be due to the big uncertainty related to the BAO measurement or the specifics of the chosen dataset. While including the Pantheon dataset decreases the deviation in general, it does not eliminate it entirely for $BAO_\theta$. This could be due to the different redshift distributions of BAO and Pantheon affecting the model fit: the maximum redshift for the binned Pantheon is $z_{max}^{SN}=1.6$ vs $z_{max}^{BA0}=2.4$ for BAO, and the median redshifts are accordingly  $\tilde{z}^{SN}=0.2$ vs $\tilde{z}^{BAO}=0.6$. Taking into consideration the big errors of the DE parameters for the different models and that all the evidences against $\Lambda$CDM are weak, we see that one needs much better BAO data to get a statistically strong preference if there is such. 

\begin{figure}
 	\centering
\includegraphics[width=0.45\textwidth]{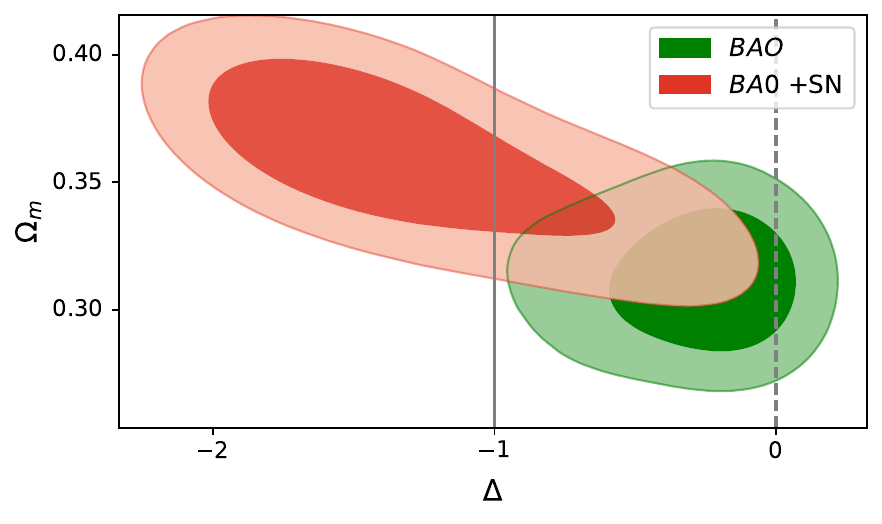}
\includegraphics[width=0.45\textwidth]{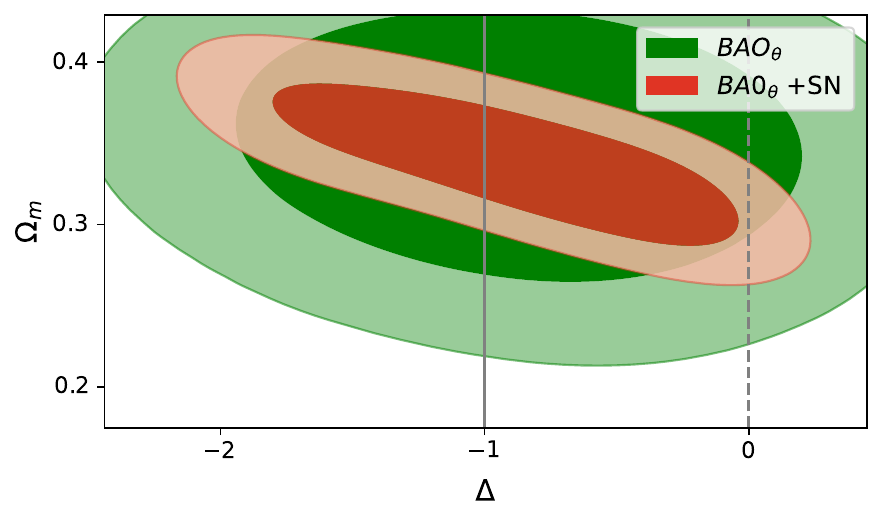}
\caption{\it{The posterior distribution for $\Omega_m$ and $\Delta$ in the gEDE model with the $BAO$ data to the upper panel and the $BAO_\theta$ data to the lower, with the solid line corresponding to pEDE and the dashed line to $\Lambda$CDM}
}
 	\label{fig:bwwa_BAO4}
\end{figure}
\begin{figure}
 	\centering
\includegraphics[width=0.45\textwidth]{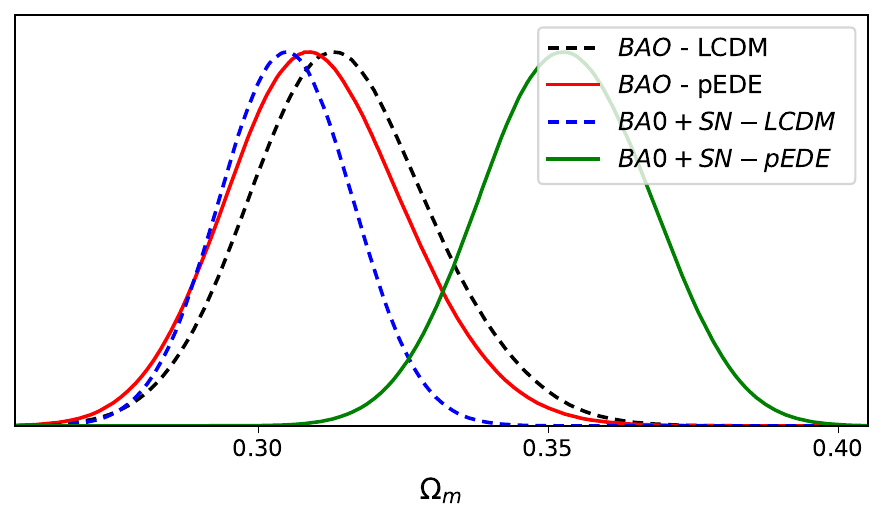}
\includegraphics[width=0.45\textwidth]{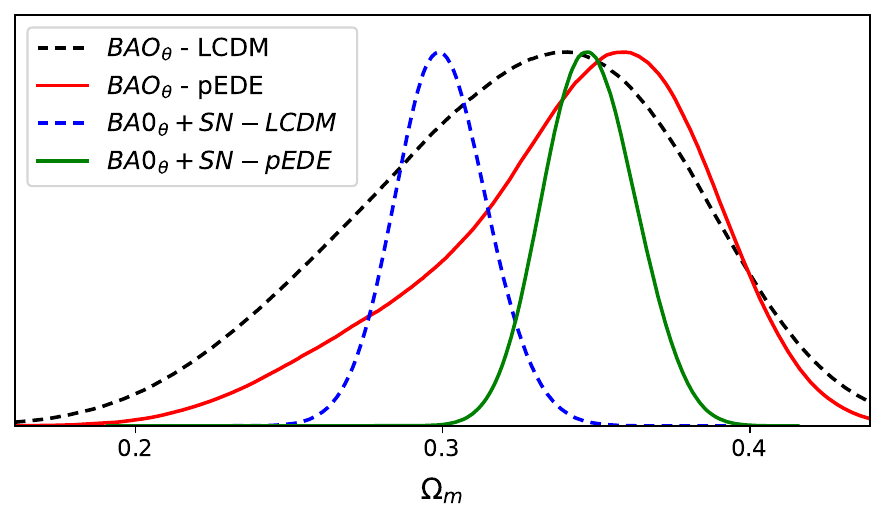}
\caption{\it{The posterior distribution for $\Omega_m$ for the two 1-parameter models: LCDM and pEDE for the $BAO$ and $BAO_\theta$ datasets}}
 	\label{fig:bwwa_BAO5}
\vspace{-0.1cm}
\end{figure}

\section{Discussion}
\label{sec:sum}

In order to avoid the problem of the degeneracy between $H_0-r_d$ in the BAO measurements, and the assumptions on the data it imposes, this paper removes the combination $H_0\cdot r_d$ entirely by marginalizing over it in the $\chi^2$. We use two different BAO datasets to test our approach. The first one -- named $BAO$ comes from different measurement provided by SDSS, WiggleZ, DES etc.,  {in additional to radial measurements coming from DR16 with their covariances}. The other dataset is the $BAO{_\theta}$ compilation measuring $\theta (z)$, which is based on angular BAO measurements obtained from analyses of luminous red galaxies, blue galaxies, and quasars. These transversal BAO data has the advantage to be weakly dependent on the cosmological model. Both  {$D_A/r_d$, $D_M/r_d$ and $D_H/r_d$} provided from the first dataset and $\theta (z)$, provided from the second one, depend only on the combination $H_0 \cdot r_d$ which we integrate out. In a similar way, one can integrate out the dependence on $H_0$ and $M_B$ in the Pantheon SnIA dataset, leaving all the the likelihoods  depending purely on the equation of state, i.e. $\Omega_m$ and the DE parameters $\Omega_\Lambda$, $w_0$ and $w_a$, which allows us to use these datasets to infer the corresponding cosmological parameters.

We find that the BAO only datasets infer very well $\Omega_m$, close to the expected values and with a small error, but they are not sufficient to constrain significantly the parameters of the DE models. The errors on $w_0$ and particularly on $w_a$ are significant within the rather wide priors we use. The errors for the $BAO{_\theta}$ dataset are larger than the errors of the $BAO$ dataset as expected.

Adding the Type Ia supernova reduces the errors, especially for the $w_0$ parameter. For the $BAO+SN$ dataset, we find $w = -0.986\pm 0.045$. For $w\,w_a$CDM we find $w_0 = -1.18\pm 0.139$, $w_a = -0.376\pm 0.672$. From the $BAO{_\theta}+SN$ dataset, we find $w = -1.08\pm 0.14$ for the wCDM model. For $w\,w_a$CDM we find $w_0 = -1.09\pm 0.09$, $w_a = -0.31\pm 0.74$. As for the curvature, $BAO+SN$ dataset prefers a closed, almost flat, universe ($\Omega_k=-0.21\pm0.07$, while $BAO{_\theta}+SN$ dataset prefers a flat one ($\Omega_k=-0.09\pm0.15$). In both cases, the gEDE model is closer to pEDE than to $\Lambda$CDM.

Comparing to the SDSS-IV results \cite{eBOSS:2020yzd}, we see that they predict $w_0 = -0.939 \pm 0.073$, $w_a= -0.31\pm 0.3$ when one considers BAO+SN+CMB, but $w_0= -0.69 \pm 0.15$ when only the BAO dataset is used. Thus our results are consistent in both cases, with the BAO+SN value for $w_0$ a little lower and the BAO only value - very close to theirs.  The mean value for $w_a$ is close, but with much larger error. But we see that in SDSS-IV results, the error on $w_a$ is also rather large. Our results also predict a negative $\Omega_k$, with larger error. One should note, however, that while we include some of the most recent BAO measurements, we include only the angular part of DR12, due to its inter-redshift covariance. Also, the $BAO_\theta$ datasets has larger inherent errors  {thus it is be expected to lead to larger errors in the inferred parameters}. Finally, under the procedure we apply, some precision is lost due to the marginalization itself. Taking into account all this, we see that the procedure we employ still gives results close to the expected. 

We perform a number of statistical tests for model comparison. The two BAO datasets show small statistical preferences for different models: $\Lambda$CDM for the $BAO$ dataset and DE (wwaCDM, but also pEDE/gEDE) for the $BAO{_\theta}$ dataset. When we add the SN dataset, $\Lambda$CDM remains the best model for $BAO+SN$ dataset, but the $BAO_{\theta}+SN$ dataset shows weak but non-negligible preference for DE models. 

Our conclusion is that one cannot constrain sufficiently the DE models from the  chosen uncalibrated, mostly angular, BAO datasets alone. Adding the Type Ia supernova to further reduce the errors and to remove some possible degeneracy helps but it only helps to constrain $w_0$ and not so much  $w_a$. However, the results on $\Omega_m$ and $w_0$ seem constrained enough to confirm the usefulness of this new approach. A downside is that for the moment, it is not possible to include all correlated $D_M-D_H$ measurements, since it is not possible to integrate out $H_0 \cdot r_d$ for a covariance matrix  {over different $z$. For this reason we have not used all known correlations in the BAO data which will improve on the errors and thus could lead to better constraints.} We predict that future measurements of the BAO would increase the efficiency of the approach as long as the correlation between some redshifts is not large. In any case, the marginalization approach offers a new perspective on the degeneracy $H_0-r_d-\Omega_m$ since in this case, the only varying parameter is $\Omega_m$ and  {it could be a tool for an independent crosscheck on DE models}.

\begin{acknowledgements}
We thank Eleonora Di-Valentino and Sunny Vagnozzi for useful comments and discussions. We would like to also thank the anonymous referee for their helpful comments regarding the manuscript. D.B. thanks to the Grants Committee of the Rothschild and the Blavatnik Cambridge Fellowships for generous supports. D.B. acknowledges a Postdoctoral Research Associateship at the Queens' College, University of Cambridge. D.B. \& D.S. is thankful to Bulgarian National Science Fund for support via research grant KP-06-N 58/5. We have received partial support from European COST actions CA15117 and CA18108. 

\end{acknowledgements}
\bibliographystyle{aa}
%
\bibliography{ref}

\appendix

\begin{figure}
 	\centering
\includegraphics[width=0.46\textwidth]{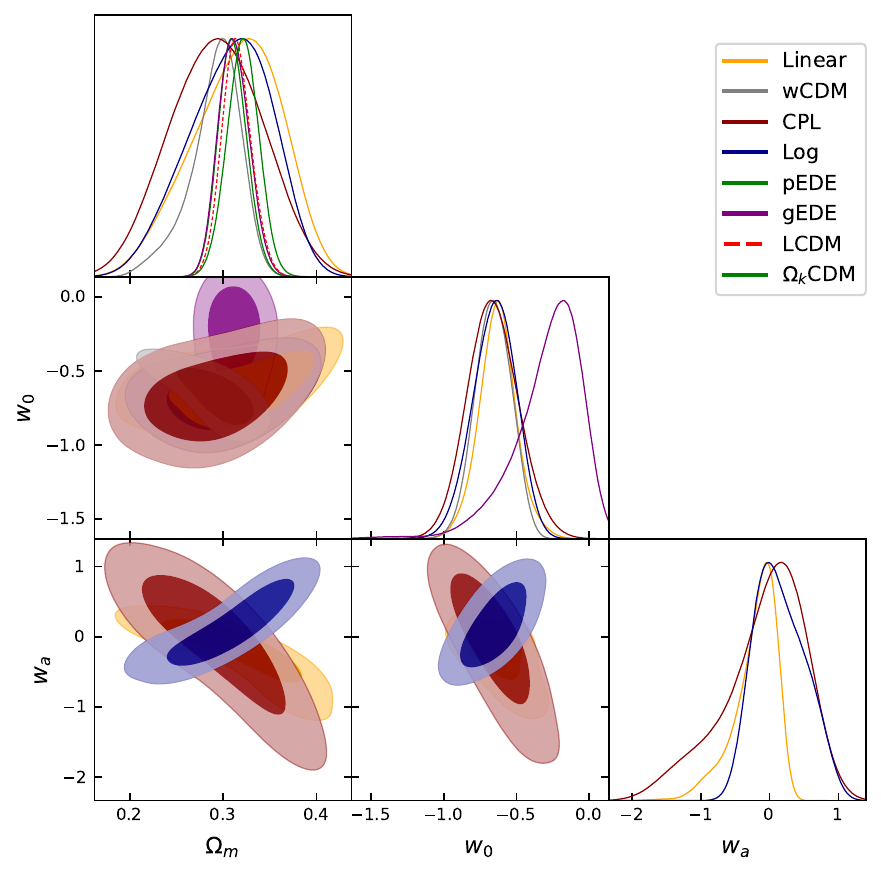}\\
\includegraphics[width=0.46\textwidth]{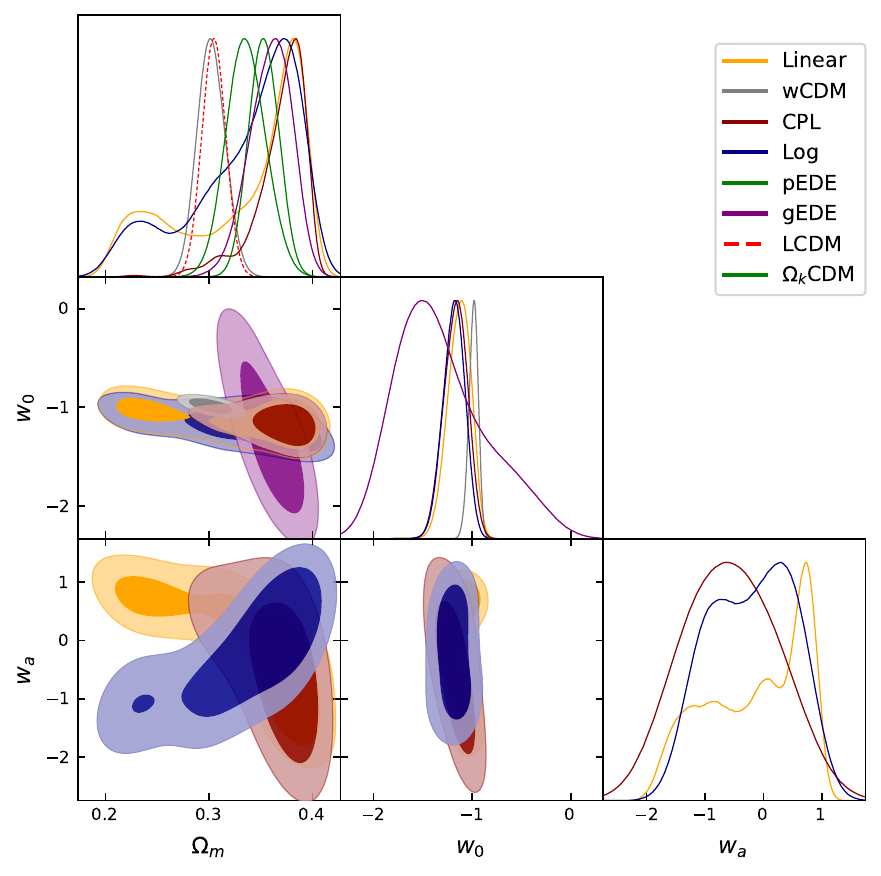}
\caption{\it{The posterior distribution for $\Omega_m$ and $w_0,w_a$ for different parametrization of DE with the $BAO$  data only on the upper and the combined BAO + Pantheon data to the lower}}
 	\label{fig:bwwa_BAO}
\end{figure}

\begin{figure}
 	\centering
\includegraphics[width=0.46\textwidth]{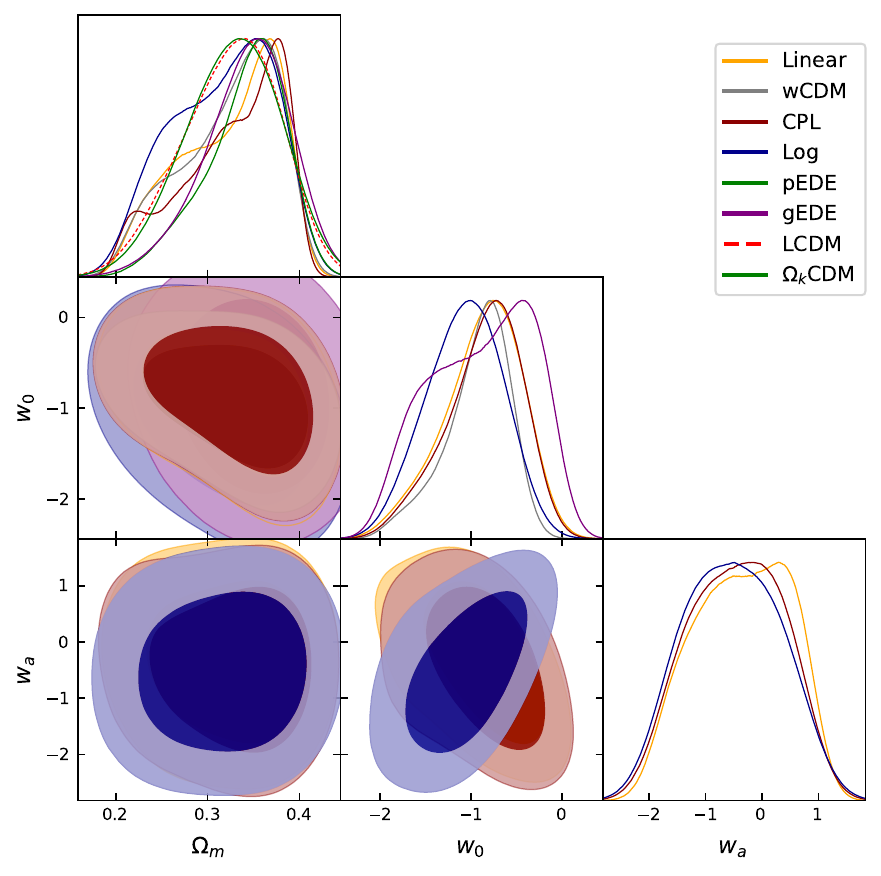}\\
\includegraphics[width=0.46\textwidth]{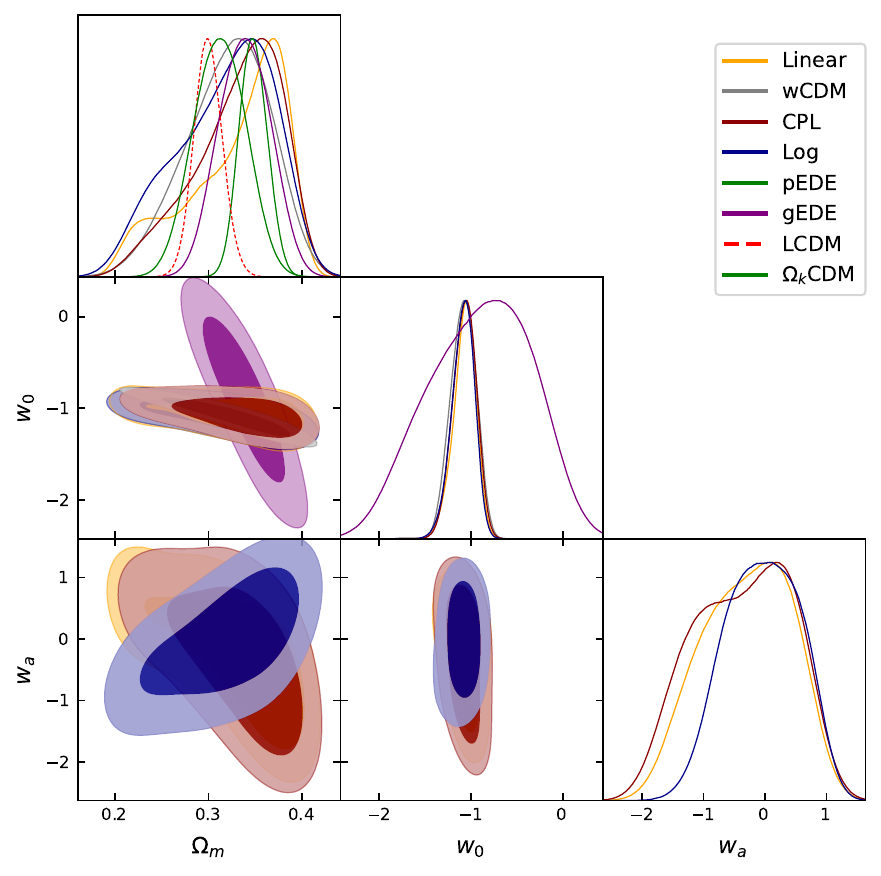}

	\caption{\it{The posterior distribution for $\Omega_m$ and $w_0,w_a$ for different parametrization of DE with the $BAO_\theta$ data only on the upper and the combined BAO + Pantheon data to the lower.}}
 	\label{fig:bwwa_SN}
\end{figure}


\end{document}